\documentclass[10pt,twocolumn]{article}
\usepackage[textwidth=7in,textheight=8.5in]{geometry}
\usepackage{graphicx}
\usepackage{amsmath} %non-Springer
\usepackage{amssymb} %non-Springer
\usepackage{fancyvrb} %non-Springer
\usepackage{soul}
\usepackage{fancyhdr}
\usepackage{epstopdf}
\usepackage{lineno}
\makeatletter
\renewcommand{\maketitle}{\bgroup
\begin{flushleft}
  \begin{Huge}
  \textbf{\@title}\\
  \end{Huge}
  \vspace{1cm}
  \@author
\end{flushleft}\egroup
}
\makeatother
\title{Study of Void Probability Scaling of Singly Charged Particles Produced in Ultrarelativistic Nuclear Collision in Fractal Scenario}
\author{%
    \textbf{{\large Susmita Bhaduri}}$^{1}$, \textbf{{\Large Dipak Ghosh}}$^{2}$\\
    $^{1}$Deepa Ghosh Research Foundation, Kolkata-700031,India \\
    $^{2}$Deepa Ghosh Research Foundation, Kolkata-700031,India \\
    \underline{$^{1}$susmita.sbhaduri@dgfoundation.in}\\
    \underline{$^{2}$deegee111@gmail.com}
}

\begin{document}
\twocolumn[
  \begin{@twocolumnfalse}
    \maketitle
  \end{@twocolumnfalse}
  ]
\noindent
%\date{}
\begin{abstract}
In this paper, we study the fractality of void probability distribution measured in $^{32}$S-Ag/Br interaction at an incident energy of $200$ GeV per nucleon. A radically different and rigorous method called \textit{Visibility Graph} analysis is used. This method is shown to reveal a strong scaling character of void probability distribution in all pseurorapidity regions. The scaling exponent, called the Power of the Scale-freeness in Visibility Graph(PSVG), a quantitative parameter related to Hurst exponent, is strongly found to be dependent on the rapidity window size. 
\end{abstract}


\section{Introduction}
\label{intro}
%10.1016/0370-1573(95)00069-0%
All through the recent decades, large density fluctuations in high energy interactions have got much consideration because of its imminent capacity to throw some light on the mechanism of multiparticle production. The presence of non-statistical fluctuations in high energy interactions has been confirmed by analyzing the fluctuations of intermittent type in both fragmentation and particle production regions \cite{Bialas1996,kit2005,dewolf1996}. 
%\cite{Ghosh2003,Sarkar2003,Bialas1996,ghosh1994,dghosh1994,Ghosh19941,dghosh1999,Jain1996}. 
Also the correlation between the emission process of produced particles and slow target fragments has been established\cite{kit2005,dewolf1996}. 
%\cite{Ghosh1982,ghosh1997,Ghosh20021}. 
Analysis of non-statistical fluctuations was believed to provide some information on the dynamics of the particle production process, so multiple methodologies have been developed to analyze non-statistical fluctuations. 
Bialas and Peschanski~\cite{bialash986} first introduced intermittency, for analysing large fluctuations. 
Scaled factorial moments of different orders showed a power law dependence on the phase space interval size in different high energy interactions. 
This was indicative of self-similarity in this type of fluctuation. 
Sarkisyan~\cite{Sarkisyan2000} has studied multiplicity distribution of the produced hadrons in high-energy interaction to describe high-order genuine correlations~\cite{Abbi1999}. In this work although the parametrization fit well the measured
fluctuations and correlations for low orders, they do show certain deviations at high orders. The necessity to incorporate the multiparticle character of the correlations along with the property of self-similarity to attain a good description of the measurements, is established~\cite{Sarkisyan2000}.
As in geometrical and statistical systems the presence of self-similarity is characteristic of fractal behavior, it has been deduced that multiparticle production process might also possess fractal characteristics and that there might be a correlation between intermittency and fractality~\cite{kit2005,dewolf1996}. 

As per Hwa and Zhang\cite{hwa2000} voids in the hadronization process is defined as region with no hardons separating the bunches of hadrons are formed during the complete course of hadronization process in a QGP system. The regions in the hadronization process, may be with high or low hadron density and no hadrons. A region without hadrons or a void is composed with quarks and gluons in a state of de-confinement in a particular time scale and they are likely to convert to hadrons a bit later. As suggested by Hwa and Zhang\cite{hwa2000} the system is divided between the state of confinement and de-confinement at the critical point and so voids exist at critical point. This concurrence of the two states at critical point is the cause of various significant behaviors of critical phenomena, so if in a heavy-ion collision, there is any phase transition, hadrons will not be produced evenly in time and space giving rise to different scaling behaviors of void distribution. Hadronic clusters in $\eta-\phi$ space might be tough to quantify, whereas voids are comparatively easy to define\cite{hwa2000}. To gain some physical understanding for the fluctuation pattern in particle production around the critical point, it is useful to analyze the production and the distribution of voids around the critical point in rapidity space in hadronic collision process. As rapidity of the particle produced, can be determined accurately, we can calculate the rapidity gaps between the pairs of adjacent particles which are essentially one dimensional representation of voids. The detailed properties of these gaps can be found in\cite{hwa2000}.

In the recent past, several studies have been done using techniques based on the fractal theory in the field of multiparticle emission\cite{hwa90,paladin1987,Grass1984,hal1986,taka1994}. Hwa(Gq moment)\cite{hwa90} and Takagi(Tq moment)\cite{taka1994} have developed the most popular of them. We have extensively applied both these methods to analyze the multipion emission process, considering their merits and pitfalls\cite{dghosh1995,dghosh2003,dghosh2002b,dghosh2002c,dghosh1995a}. Thereafter methodologies like \textit{Detrended Fluctuation Analysis (DFA)} method\cite{cpeng1994} has been used for determining \textit{mono}fractal scaling exponents and for recognizing long-range correlations in noisy and non-stationary time series data\cite{MSTaqqu1995,zchen2002}. DFA method has been extended by Kantelhardt et al.\cite{kantel2002}, to analyze non-stationary and \textit{multi}fractal time series. This generalized DFA is termed as the \textit{multi}fractal-DFA (\textit{MF-DFA}) method. 
 
The DFA function obeys a power-law relationship with the scale parameter of a time series, if it is long-range correlated. If the DFA function of the time series is, say denoted by $F(s)$, it will vary with a power of its scale parameter, say denoted by $s$, by following the equation $F(s)\propto s^H$. The exponent $H$ in this equation is termed as \textit{Hurst exponent}. $H$, is related to the fractal dimension denoted by $D_F$, as $D_F=2-H$\cite{kantel2001}. Then DFA method is extended to MF-DFA technique\cite{kantel2002} which is used for this kind of analysis for its highest precision in the scaling analysis. Hurst exponent and MF-DFA parameter are used widely in nonlinear, non-stationary analysis and they have accomplished in identifying long-range correlations for different time series. 
Zhang et al.\cite{YXZhang2007} applied MF-DFA method to analyze the multifractal structure of the distribution of shower tracks around central rapidity region of Au-Au collisions at $\sqrt{s}_{NN}=200$A GeV. Multifractal analysis in particle production process has been done in various works of recent times~\cite{Albajar1992,Suleymanov2003,Ferreiro2012}.
But the DFA method is constrained by the requirement of an \textit{infinite} number of data points for the time series to give most accurate value of the Hurst exponent. 
Whereas, in a real-life scenario, for most of the times, getting infinite number of data points is not possible. So we have to use \textit{finite} time series for calculation of the Hurst exponent. In this process the long-range correlations in the time series are broken in parts with finite number of data points and consequently the local dynamics relating to a specific temporal window are evidently amplified and diverges from its correct form.

In this regard more accurate result may be obtained with an entirely different, rigorous method namely the \textit{Visibility graph analysis}\cite{laca2008,laca2009}. Lacasa et al. have used fractional Brownian motion(fBm) and fractional Gaussian noises(fGn) series as a theoretical framework to study real-time series in various scientific fields. Due to the inherent non-stationarity and long-range dependence in fBm, the Hurst parameter calculated for it with various methods, often yields ambiguous results. Lacasa et al. showed how classic method of complex network analysis can be applied to quantify long-range dependence and fractality of a time series\cite{laca2009}. He mapped fBm and fGn series into a scale-free Visibility graph having the degree distribution as a function of the Hurst exponent\cite{laca2009}.
Visibility Graph analysis is altogether a new concept for assessing fractality from a new perspective without assessing multifractality. 
This method has recently been applied widely over time series with \textit{finite} number of data points, even with $400$ data points\cite{jiang2013}, and has achieved reliable result in various fields of science.
In our recent works we have applied Visibility graph analysis productively for analyzing various biological signals~\cite{Bhaduri2014,Bhaduri2016}.
Zebende et al. have studied long-range correlations because of temperature-driven liquid to vapor phase transition in distilled water, using DFA method and have established that with the temperature approaching transition temperature, the scaling exponent increases~\cite{zebende2004}. Recently Zhao et al. has confirmed that Hurst exponent may be a good indicator of phase transition for a complex system~\cite{zhao2016}.
%Recently Zhao et al. has confirmed that Hurst exponent may be a good indicator of phase transition for a complex system[Ref.~arXiv:1601.07715].
We also know from his analysis on magnetization time series with Hurst exponent, that as the system approaches towards phase transition the fractal behavior of the system transforms from mono to multifractal. 
Zhao et al. have applied Visibility network analysis to confirm the fractality of the same time series, from complex network perspective~\cite{zhao2016} and reported that remarkable increase or decrease of the topological behavior can be used to identify the onset of phase transition.
Lacasa et al. have shown that the Visibility graph method is appropriate for distinguishing correlation, geometrical structure of a time series and have established a connection between the Hurst exponent and the Power of Scale-freeness in Visibility Graph(\textit{PSVG})~\cite{laca2009}.

In case of high energy interactions, void occurrence probability can be defined as the probability of occurring events with zero number of particles in a particular region of phase space.
So far no work has been reported on scaling analysis of void probability distribution of pions with respect to fractal and multifractal methodology.  
We have analyzed void probability distribution data for different windows of pseudorapidity around the central rapidity region. 
In this work we have attempted to analyze the scaling behavior of voids in multipion production from a completely new perspective of Visibility Graph analysis as well as to explore phase transition if any, in $^{32}$S-Ag/Br interaction at an incident energy of $200$ GeV per nucleon. We have analyzed the data in different overlapping windows, ranging from central rapidity region to the full phase space.

The rest of the paper is organized as follows. The method of Visibility graph technique is presented in Section~\ref{ana}. The details of data, our analysis and the inferences from the test results are given in Section~\ref{exp}. The paper is concluded in Section~\ref{con}.

\section{Method of analysis}
\label{ana}

As discussed earlier the \textit{Visibility graph technique} can be broadly used over finite time-series data set and can produce reliable results in several domains. The simple method converts a fractal time series into a scale-free graph, and its structure is related to a self-similar fractal nature and complexity of the time series\cite{laca2008}. In this technique each node of the scale-free graph which is also called the visibility graph represents a time sample of the time series, and an edge between two nodes shows the corresponding time samples that can view each other. 
\subsection{Visibility Graph Algorithm}
\begin{figure}
\includegraphics[scale=.4]{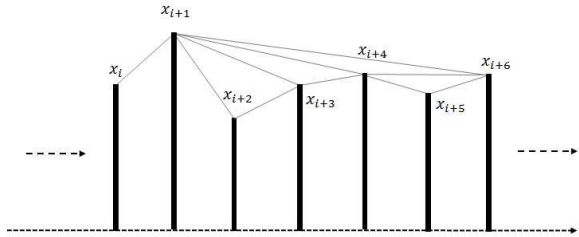}
\caption{Visibility Graph for time series X. \label{visi}}
\end{figure}
The algorithm is a one-to-one mapping from the domain of time series $X$ to its Visibility Graph. Let $X_i$ be the $i^{th}$ point of the time series. $X_{m}$ and $X_{n}$ are the two vertices(nodes) of the graph, that are connected via a bidirectional edge if and only if the following equation is valid, where $\forall j \in Z^{+} \mbox{ and } j < (n-m)$.
\setcounter{equation}{0}
\begin{equation}
X_{m+j} < X_{n} + (\frac{n - (m+j)}{n-m})\cdot(X_m - X_n) 
\end{equation}
As shown in Fig.\ref{visi}, $X_m$ and $X_n$ can see each other, if the above is satisfied. With this logic two sequential points of the time series can always see each other hence all sequential nodes are connected together. The time series should be converted to positive planes as the above algorithm is valid only for positive $x$ values in the time series. 

\subsubsection{Power of Scale-freeness of VG \mbox{-} PSVG}
\label{psvg}
As per the graph theory, the definition of the degree of a node is the number of connections or edges that the node has with other nodes. The degree distribution, say $P(k)$, of a network formed from the time series, is defined as the fraction of nodes with degree $k$ in the network. Thus if there are $n$ nodes in total in a network and $n_k$ of them have degree $k$, then $P(k) = n_k/n$.

The scale-freeness property of Visibility Graph states that the degree distribution of its nodes satisfies a \textit{Power Law}, i.e. $P(k) \sim k^{-\lambda_p}$, where $\lambda_p$ is a constant and it is known as \textit{Power of the Scale-freeness in Visibility Graph-PSVG}, which is denoted by $\lambda_p$ and is calculated as the gradient of $log_2[P(k)]$ versus $log_2[1/k]$ plot. $\lambda_p$ corresponds to the amount of complexity and fractal nature of the time series indicating the \textit{Fractal Dimension} of the signal
\cite{laca2008,laca2009,meh2012}. It is also proved that there exists a linear relationship between $\lambda_p$ and $H$ of the associated time series\cite{laca2009}.

\section{Experimental Details}
\label{exp}
\subsection{Data Description}

For this experiment, the data were acquired by exposing Illford G5 emulsion plates to a $^{32}$S-beam of $200$ GeV incident energy per nucleon, from CERN. The details of the data including scanning, measurement, resolution etc. were given in some of our previous works.

In this experiment, the emission angle-$\eta$ and azimuthal angle-$\phi$ are measured for each track with respect to the
beam directions. The readings are taken for the coordinates-$(X_0,Y_0,Z_0)$ of the interaction point, coordinates-
$(X_1,Y_1,Z_1)$ of the end of the linear portion of every secondary track and coordinates-$(X_i,Y_i,Z_i)$ of a point on the incident beam. The co-ordinates are measured by semi-automatic measurement system with precision of $1\mu m$ along $X$-axis and $Y$-axis. The corresponding precision in $Z$-direction is $0.5\mu m$.
In spite of its several limitations, nuclear emulsion experiments are superior than many experiments because they offer a very high angular resolution(around $1mrad$). This advantage is relevant where the distribution of particles is confined into a small phase space region. As detector it also has the ability to register all the charged particles produced or emitted in the $4\pi$- geometry space.

\subsection{Our Method of Analysis}
To analyze the void probability distribution data of produced pions, we have divided the overall $\eta$-distribution of singly charged particles produced in $^{32}$S-Ag/Br interaction at $200$A GeV, in $8$ overlapping pseudorapidity(denoted by $\eta$) windows around the central rapidity region. Below are the details of the analysis.

\begin{itemize}
\item $8$ pseudorapidity windows(denoted by $\Delta\eta$) of pseudorapidity values around the central rapidity regions were obtained. 
Hence $\eta$-values in each dataset ranges from ($c_r-(\Delta\eta)$) to ($c_r+(\Delta\eta)$), where $c_r$ is the central rapidity for the whole dataset.
The $\Delta\eta$ ranges from $\Delta\eta=0.5$ which is the narrowest region around central rapidity, to the full phase space, $\Delta\eta=4$.  

\item Each dataset of $\eta$-values are sorted in ascending order, and this way $8$ series of data points are formed. Then for each data series gaps or voids between the $\eta$-values are noted. Then, starting from minimum $\eta$-value to maximum $\eta$-values, the series is divided into a number of windows with a specific width. Then for each window, we calculate the number of voids and divide it by the total number of voids for the whole data series. Thus we get the void probability distribution for the particular data series. Each dataset of void probability distribution contains number of points equal to the number of windows that could be made from the data series of $\eta$-values. Visibility Graph analysis is done on these datasets of void probability distribution.

\item Then Visibility Graph is constructed following the method in Sec.~\ref{ana}, for each of the $8$ datasets(consisting of void probability values) generated out of overlapping rapidity intervals around the central rapidity region. For each Visibility Graph the values of $k$ vs $P(k)$ are calculated. 

The $k$ vs $P(k)$ plot dataset for void probability distribution for one of the overlapping windows of $\eta$-values for $^{32}$S-Ag/Br interaction is shown in Fig~\ref{power}, and the power law relationship is obvious here.

\begin{figure}
\centerline{\includegraphics[width=3.5in]{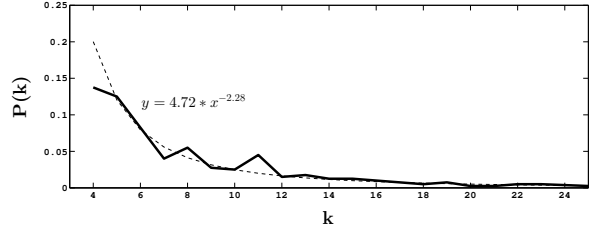}}
\caption{$k$ vs $P(k)$ plot for the Visibility Graph created for a test dataset of void probability distribution. \label{power}}
\end{figure}

\item As per the method described in Section.~\ref{ana}, $\lambda_p$ is deduced from the slope of $log_{2}[1/k]$ versus $log_{2}[P(k)]$ for each of the $k$ vs $P(k)$ dataset. 

\begin{figure}
\centerline{\includegraphics[width=3.5in]{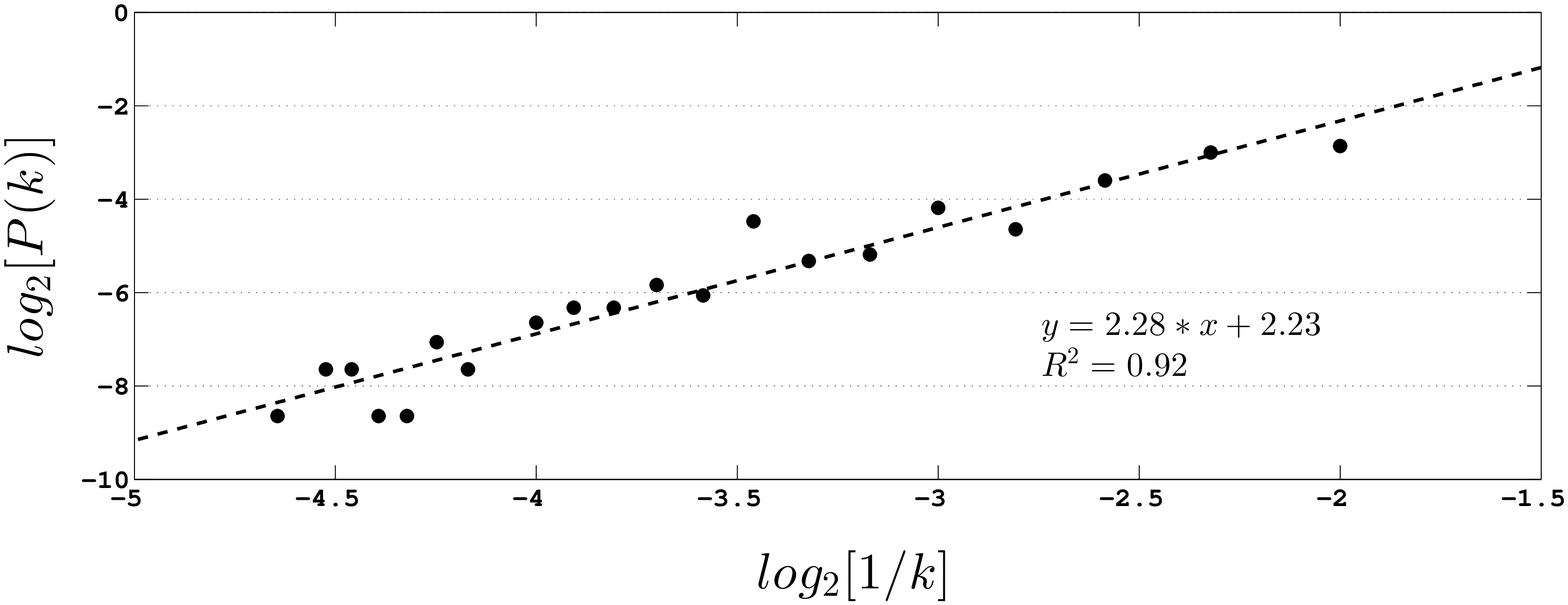}}
\caption{Slope of $log_{2}[1/k]$ versus $log_{2}[P(k)]$ for the Visibility Graph created for the same dataset in Fig~\ref{power}. \label{fit}}
\end{figure}

Fig~\ref{fit} shows the plot of $log_{2}[1/k]$ versus $log_{2}[P(k)]$ for the same dataset for which the $k$ vs $P(k)$ plot is shown. It is evident from the Fig~\ref{fit} that the value of $\lambda_{p}$, calculated for this dataset is $2.28\pm0.11$.

\item Table.~\ref{result} lists $\lambda_p$-values(along with the statistical errors), denoted by $\lambda_{p^{exp}}$, for $8$ sets of void probability dataset for the $^{32}$S-Ag/Br interaction. Similar analysis using randomized data are also included.
Further for the sake of comparison we have included another Table~\ref{resulteta} where similar analysis were performed on pseudorapidity($\eta$) values.

\begin{table}[h]
\centering
\caption{Values of $\lambda_{p^{exp}}$ for experiment data and $\lambda_{p^{rand}}$ for randomized data, calculated for void probability distribution for different $\eta$-range around the central rapidity regions.} 
\label{result}
\begin{tabular}{|c|c|c|}
\hline
\textbf{$\Delta\eta$}&\textbf{$\lambda_{p^{exp}}$}&\textbf{$\lambda_{p^{rand}}$}\\
\hline
$0.5$&$1.87\pm0.15$&$2.66\pm0.50$\\
1.0&$1.96\pm0.15$&$3.19\pm0.20$\\
1.5&$2.37\pm0.15$&$3.82\pm0.30$\\
2.0&$2.28\pm0.11$&$2.66\pm0.27$\\
2.5&$2.33\pm0.09$&$2.68\pm0.17$\\
3.0&$2.18\pm0.07$&$2.52\pm0.18$\\
3.5&$2.15\pm0.07$&$2.90\pm0.21$\\
4.0&$1.99\pm0.05$&$2.55\pm0.19$\\
\hline
\end{tabular}
\end{table}

\begin{table}[h]
\centering
\caption{Values of $\lambda_{p}$ calculated for multipion production in $^{32}$S-Ag/Br interactions, $\eta$-range wise around the central rapidity regions}
\label{resulteta}
\begin{tabular}{|c|c|}
\hline
\textbf{$\Delta\eta$}&\textbf{$\lambda_{p^{exp\eta}}$}\\
\hline
$0.5$&$1.15\pm0.06$\\
$1.0$&$1.46\pm0.05$\\
$1.5$&$1.26\pm0.03$\\
$2.0$&$1.13\pm0.02$\\
$2.5$&$1.02\pm0.02$\\
$3.0$&$0.98\pm0.02$\\
$3.5$&$0.99\pm0.02$\\
$4.0$&$0.99\pm0.02$\\
\hline
\end{tabular}
\end{table}

\end{itemize}

\section{Conclusion} 
\label{con}
\begin{itemize}
\item This analysis with chaos-based complex network analysis clearly manifests that the void probability distribution in multiparticle production in relativistic nuclear collision obeys scaling behavior.

\item Further compared to randomized data it is observed that, this scaling behavior depends on the rapidity window size in and around the central pseudorapidity region - although the dependency becomes weak when statistical errors are considered. However the overall picture speaks in favor of interval size dependence of void probability distribution which is a new finding obtained from a rigorous, complex, nonlinear technique. 

\item This sort of analysis has a potential for assessment of approaching criticality for phase transition in terms of PSVG values, as also discussed in~\cite{zebende2004,zhao2016}.

\end{itemize}

\section{Acknowledgement} 
\label{ack}
We thank the \textit{Department of Higher Education, Govt. of West Bengal, India} for logistics support of computational analysis.

\section{Disclosure} 
\label{dis}
The author(s) declare(s) that there is no conflict of interest regarding the publication of this manuscript.

\section{References}
\bibliographystyle{splncs} 
\bibliography{partvoid}
\end{document}